\documentclass[conference,10pt]{IEEEtran}
 \IEEEoverridecommandlockouts
\usepackage{cite}
\usepackage{color}
\ifCLASSINFOpdf
	\usepackage[pdftex]{graphicx}
	\graphicspath{{./figures/}}
\else
	\usepackage[dvips]{graphicx}
	\graphicspath{./figures/}
\fi
\graphicspath{{fig/}{jpeg/}}
\usepackage[cmex10]{amsmath}
\usepackage{amssymb}
\usepackage{algorithm}
\usepackage{algorithmic}
\usepackage{subcaption}
\usepackage{caption}
\usepackage[outdir=./]{epstopdf}
\usepackage{comment}
\usepackage{bm}
\input{mysymbol.sty}
\usepackage{needspace}

\usepackage{tikz}
\usetikzlibrary{shapes,arrows}
\usepackage{float,environ}
\usetikzlibrary{matrix} 
\usetikzlibrary{decorations.markings} 
\usetikzlibrary{calc} 
\usetikzlibrary{decorations.pathmorphing,snakes, positioning, decorations.pathreplacing}

\usetikzlibrary{fit}

\usepackage{pgfplots}
\usepackage{color}
\usepackage{multirow}
\usepackage{setspace}
\makeatletter
\newsavebox{\measure@tikzpicture}
\NewEnviron{scaletikzpicturetowidth}[1]{%
  \def\tikz@width{#1}%
  \begin{lrbox}{\measure@tikzpicture}%
  \BODY
  \end{lrbox}%
  \pgfmathparse{#1/\wd\measure@tikzpicture}%
  \BODY
}
\makeatother

\newcommand{\QED}{\hfill\ensuremath{\blacksquare}}

\def\vec2{\text{vec}}

\newtheorem{theorem}{\hspace{0pt}\bf Theorem}

\begin{document}
%
\title{Resource Allocation via Graph Neural Networks in Free Space Optical Fronthaul Networks \thanks{Supported by ARL DCIST CRA W911NF-17-2-0181 and Intel Science and Technology Center for Wireless Autonomous Systems.}
}

\author{\IEEEauthorblockN{Zhan Gao$^{*}$ \qquad Mark Eisen$^{\dagger}$  \qquad Alejandro Ribeiro$^{*}$ }
\IEEEauthorblockA{$^{*}$Department of Electrical and System Engineering, University of Pennsylvania, Philadelphia, PA, USA \\
$^{\dagger}$Intel Corporation, Hillsboro, OR, USA  \\
Email: gaozhan@seas.upenn.edu, mark.eisen@intel.com, aribeiro@seas.upenn.edu}
}

\maketitle

\begin{abstract}
This paper investigates the optimal resource allocation in free space optical (FSO) fronthaul networks. The optimal allocation maximizes an average weighted sum-capacity subject to power limitation and data congestion constraints. Both adaptive power assignment and node selection are considered based on the instantaneous channel state information (CSI) of the links. By parameterizing the resource allocation policy, we formulate the problem as an unsupervised statistical learning problem. We consider the graph neural network (GNN) for the policy parameterization to exploit the FSO network structure with small-scale training parameters. The GNN is shown to retain the permutation equivariance that matches with the permutation equivariance of resource allocation policy in networks. The primal-dual learning algorithm is developed to train the GNN in a model-free manner, where the knowledge of system models is not required. Numerical simulations present the strong performance of the GNN relative to a baseline policy with equal power assignment and random node selection.
\end{abstract}


\section{Introduction}

5G wireless networks are expected to provide high rates, low latency and flexible constructions through an ultra-dense deployment of small cells \cite{Gupta2015}. The cloud radio access network (C-RAN) emerges as a promising cellular architecture to satisfy these requirements \cite{Wu2015}. It moves the baseband signal processing to a centralized baseband unit (BBU) pool, and distributed remote radio heads (RRHs) are responsible for capturing signals and forwarding them to the BBU. RRHs are connected to the BBU via fronthaul links. These links are traditionally optical fibers with high capacity and zero interference \cite{Poor2015}. However, optical fiber is not ubiquitous and its deployment can be expensive. Free space optical (FSO) communication provides an attractive alternative that maintains comparable advantages as optical fiber \cite{Alzenad2018, Ahmed2018}. More importantly, it is cost-efficient and flexible in implementation. On the other hand, FSO links are sensitive to channel characteristics and may be significantly impacted by factors like weather and turbulence. Various models are put forth to characterize the FSO channel, based on which different techniques are developed to reduce channel effects \cite{Andrews2005, Gao2017, Gao2018, zhang2019ergodicity}.

To mitigate the channel dependent degradation, cooperative transmissions has been proposed in FSO networks. With a certain transmit power budget, adaptive powers are assigned based on the CSI to maximize the channel capacity \cite{ Zhou2013, Zhou2015}. For the relayed communication, joint relay selection and power adaptation are proposed to minimize the outage probability or maximize the network throughout \cite{Chatzidiamantis2013, Hassan2018, Hassan2017}. However, the aforementioned works employ relaxations to simplify problem models leading to approximate solutions, and are computationally expensive to implement with respect to instantaneous channel state changes. These issues motivate the application of machine learning due to its low complexity and potential for model-free implementation. Deep learning in particular has been leveraged for resource allocation problems in wireless radio frequency domain \cite{Sun2018, Xu2017, Eisen2019}. In \cite{Gao2019}, we proposed the deep neural network to solve the optimal power allocation for WDM radio on free space optical (RoFSO) systems. Deep neural networks are limited, however, in their ability to generalize over varying network topologies.

As an extension of convolutional neural networks, graph neural networks (GNNs) have been used to model and analyze data collected from networks and has achieved signifigant successes in many learning tasks \cite{Henaff2015, Gama2019, Eisen2020, Zhan2020}. With the use of graph convolutional filters, GNNs exploit the underlying irregular structure of network data in a manner that results in lower computation complexity, less parameters for training, and generalization capabilities relative to traditional deep neural networks. As such, the GNN is considered as a suitable candidate learning over network scenarios such as the FSO fronthaul network and related communication networks.

In this paper, we consider the resource allocation problem in the FSO fronthaul network of C-RAN architecture. RRHs transmit optical signals to intermediate aggregation nodes (ANs) through FSO links, and the latter forward aggregated signals to the BBU through high speed fiber. Based on the CSI, different powers are assigned to different RRHs and an optimal AN is selected at each RRH to maximize the objective function, subject to power limitation and data congestion constraints. The formulated optimization problem is non-convex with complicated constraints (Section II). We introduce the graph neural network (GNN) to parameterize the resource allocation policy and translate the problem into a statistical learning problem (Section III). The GNN is shown to be permutation equivariant in the sense that relabelling RRHs or ANs in network results in the same optimal policy. The primal-dual learning algorithm is developed to train the GNN solving the learning problem (Section IV). With the use of policy gradient, its implementation is completely model-free that avoids the error induced by the system model inaccuracy. Numerical simulations are present to show significant performance of the proposed GNN learning algorithm (Section V).



\section{Problem Formulation}\label{sec_problem}

\begin{figure}[t]
\centering
\includegraphics[width=0.9\linewidth , height=0.45\linewidth, trim=20 20 20 20]{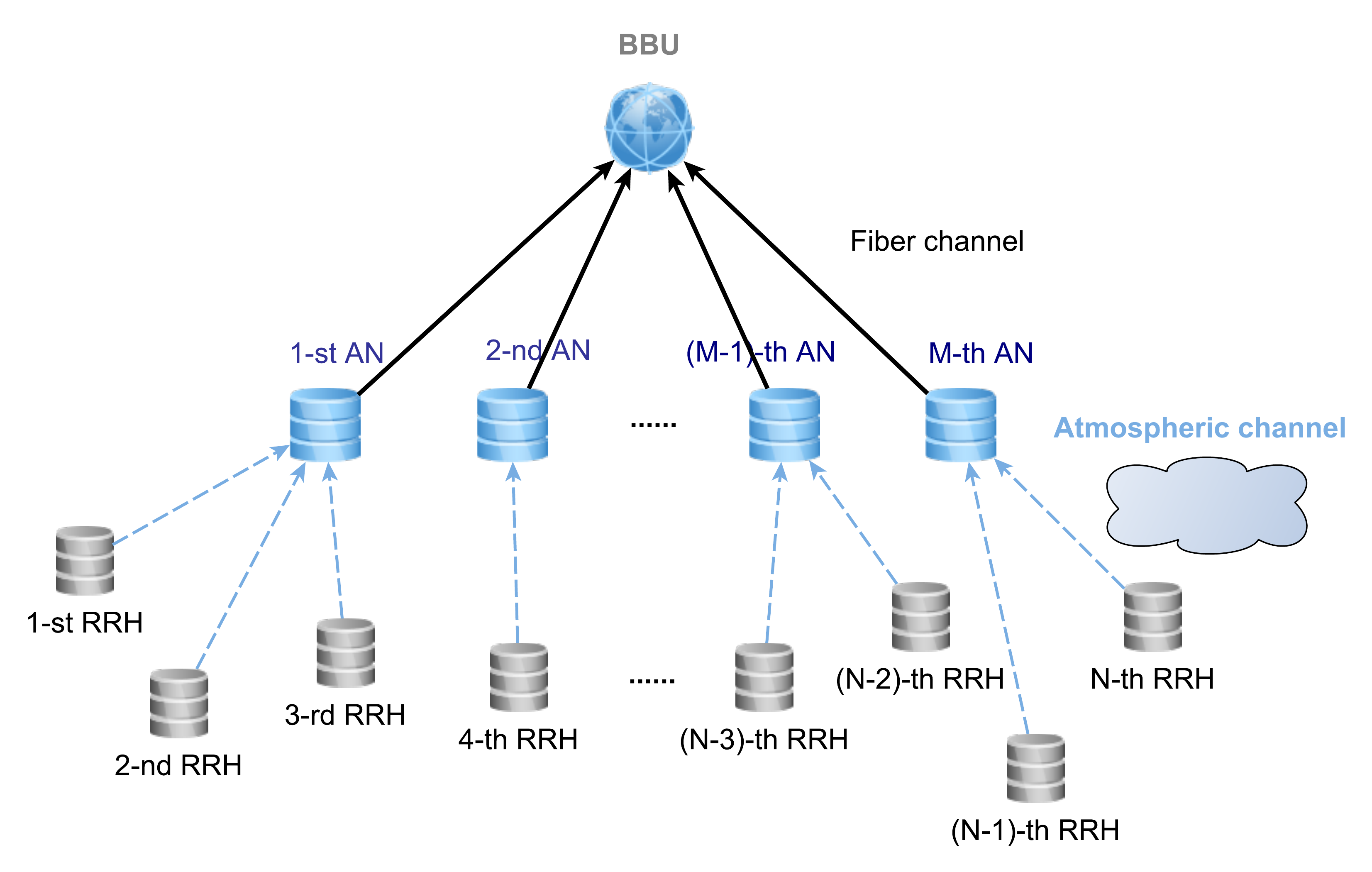}
\caption{The FSO fronthaul network. }
\label{fig_network}
\end{figure}

We consider the free space optical fronthaul network consists of remote radio heads, aggregation nodes and the baseband unit. RRHs transmit optical signals through free space to selected ANs. ANs then aggregate received signals and forward them to the BBU through optical fiber. The RRHs are distributed remotely in the space around ANs and equipped with multiple apertures pointing towards ANs, such that each RRH can possibly transmit signals to all ANs. The ultra-wide field of view (UWFOV) beam acquisition is available in ANs, through which each AN can receive optical signals from different RRHs simultaneously. See Fig. 1 for details of the proposed FSO fronthaul network. The resource allocation problem considered here comprises the power adaptation and the AN selection. Based on the CSI, different powers are assigned to different RRHs and each RRH selects a best AN for signal transmission, in order to maximize the objective function. The exact objective can be adjusted according to practical situations.

Assume there are $N$ RRHs, $M$ ANs and one BBU in a FSO fronthaul network. The CSI between RRHs and ANs is represented by the matrix $\bbH \in \mathbb{R}^{N \times M}$, where $[\bbH]_{nm} = h_{nm}$ is the CSI between $n$-th RRH and $m$-th AN for all $n=1,\ldots,N$ and $m=1,\ldots,M$. The assigned power and the selected AN for $n$-th RRH is based on the observed CSI $\bbH$ via a power assignment policy $P_n(\bbH)$ and an AN selection policy $\bbalpha_n(\bbH)=[\alpha_{n1}(\bbH),\ldots,\alpha_{nM}(\bbH)]^\top$. Here $\alpha_{nm}(\bbH) \in \{ 0,1 \}$ is the indicator taking one if $m$-th AN is selected by $n$-th RRH and zero if not, such that $\sum_{m=1}^M \alpha_{nm}(\bbH) \le 1$. In addition to channel states, separate state variables $\bbx \in \mathbb{R}^{N+M}$ represent the status of RRHs and ANs, such as equipment conditions of RRHs and ANs. Given the collection of resource allocations $\bbP(\bbH)= [P_1(\bbH),\ldots,P_N(\bbH)]$ and $\bbDelta(\bbH)=[\bbalpha_1(\bbH),\ldots,\bbalpha_M(\bbH)]$ with network states $\bbH$ and $\bbx$, a channel capacity $C_{nm}(\bbx, \bbH,\bbP(\bbH), \bbDelta(\bbH))$ is achieved between $n$-th RRH and $m$-th AN. Since the FSO channel is a fading process with coherence time on the order of milliseconds, we consider it as an ergodic and i.i.d block fading process. The instantaneous CSI tends to vary fast, so as the instantaneous channel capacity. Therefore, a long term average $\mathbb{E}\left[ C_{nm}(\bbH,\bbP(\bbH)) \right]$ is the more meaningful metric. We then consider the objective function as the weighted sum of channel capacities over RRHs
\begin{equation}
\sum_{n=1}^N  \omega_n \sum_{m=1}^M \mathbb{E}\left[ C_{nm}(\bbx,\bbH,\bbP(\bbH),\bbDelta(\bbH)) \right],
\end{equation}
where $\bbomega = [\omega_1,\ldots,\omega_N]^\top$ is the weight vector representing priorities of different RRHs and the expectation $\mathbb{E}[\cdot]$ is with respect to the probability distribution of the CSI $\bbH$.

Three constraints are considered. Assuming RRHs are connected to a common power supply, the first is the expected total power limitation
\begin{equation}
\mathbb{E} \left[ \sum_{n=1}^N P_n(\bbH) \right] \le P_t.
\end{equation}
The second is motivated by the safety concern. To avoid possible danger in optical beam transmission, we limit the peak power that can be allocated on single optical signal
\begin{equation}\label{eq:safe}
0 \le P_n(\bbH) \le P_s,~\forall n=1,...,N.
\end{equation}
The third is to prevent the data congestion at ANs. Specifically, we require the incoming sum-capacity at each AN less than or equal to the capacity of optical fiber
\begin{equation}
\sum_{n=1}^N \mathbb{E}\left[ C_{nm}(\bbH,\bbP(\bbH),\bbDelta(\bbH)) \right] \le C_t,~\forall m=1,\ldots,M.
\end{equation}
Together, we formulate the resource allocation problem in the FSO fronthaul network as follows
\begin{alignat}{3} \label{eq_problem111}
 \mathbb{P}&:=   \max_{\bbP, \bbDelta} \  \sum_{n=1}^N  \omega_n \sum_{m=1}^M \mathbb{E}\left[ C_{nm}(\bbx,\bbH,\bbP(\bbH),\bbDelta(\bbH)) \right] ,             \\
        &  \st \quad \mathbb{E} \left[ \sum_{n=1}^N P_n(\bbH) \right] \le P_t,~ 0 \le P_n(\bbH) \!\le\! P_s,  \nonumber \\
        &  \quad \quad\quad  \sum_{m=1}^M \!\alpha_{nm}(\bbH)\leq1,\forall n=1,...,N,   \nonumber   \\
       &       \quad\!\! \sum_{n\!=\!1}^N\! \mathbb{E}\left[ C_{nm}(\bbx,\bbH,\bbP(\bbH),\bbDelta(\bbH)) \right] \!\le\! C_t,\forall m\!=\!1,...,M.   \nonumber
\end{alignat}
While some theoretical models exist for the channel distribution, e.g. \cite{Andrews2005}, and the channel capacity for FSO fronthaul networks, e.g. \cite{Hassan2017}, we stress that the above problem is given without any particular system model. Pre-existing models may not be accurate in practical systems, leading to inevitable errors for model-based algorithms. We therefore propose a ``model-free'' approach in this paper in which no explicit knowledge or form of any of the models in \eqref{eq_problem111} is assumed.


\section{Graph Neural Networks}\label{sec_gnn}

\begin{figure}
\centering
\includegraphics[width=0.35\textwidth, trim=0 0 0 0]{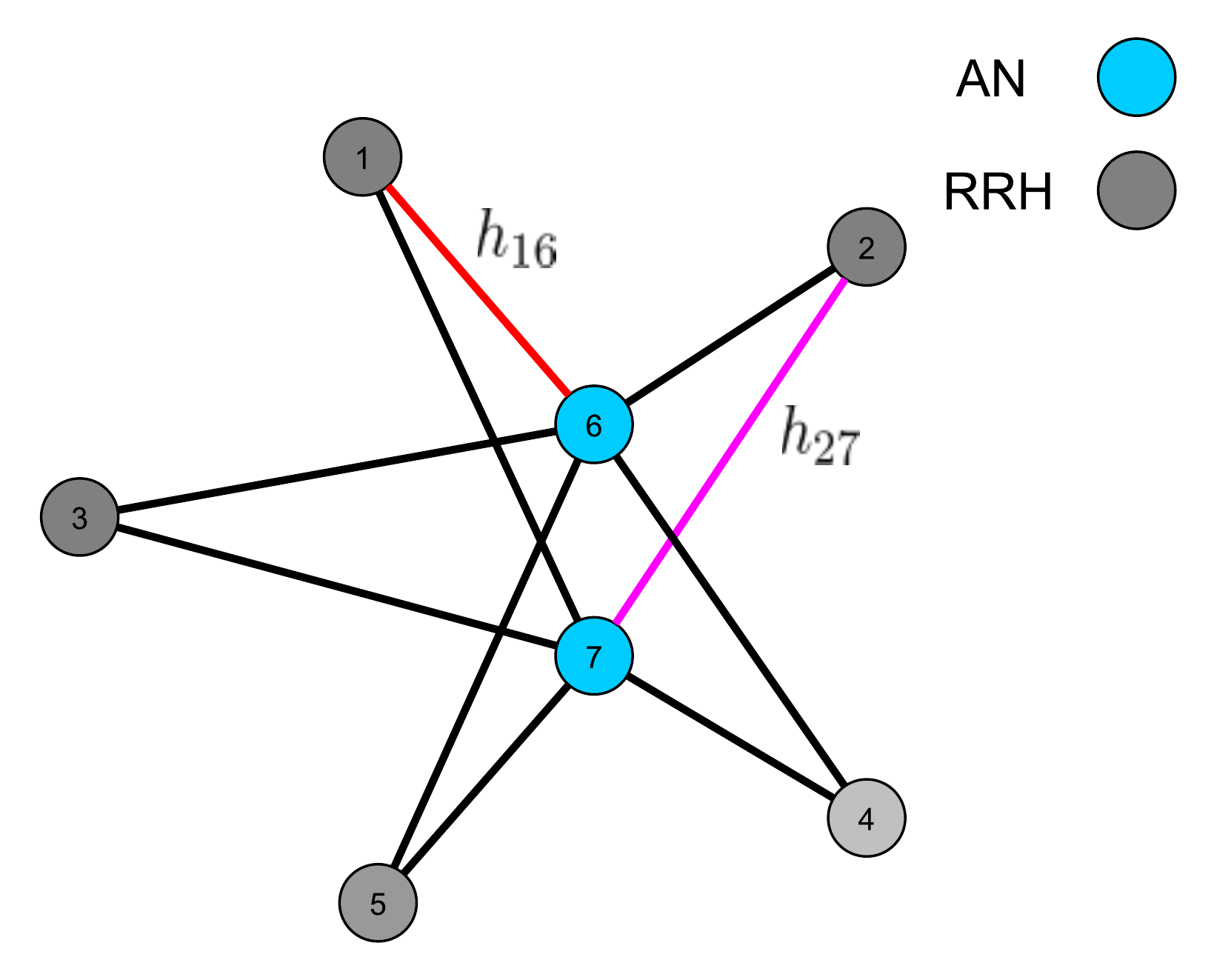}
\caption{The weighted bipartite graph representation of an FSO fronthaul network with $N=5$ RRHs and $M=2$ ANs. The RRHs (grey nodes) are connected only to ANs (blue nodes) and vice versa through FSO links. The edge weight between nodes is the CSI of the associated link.}
\label{fig_gnn}
\end{figure}

With unknown objective function and complicated constraints, the optimization problem \eqref{eq_problem111} is challenging. We pursue to solve it with a learning procedure. In particular, the resource allocation policies are functions of the CSI $\bbH$. By introducing a parameterization $\bbtheta \in \mathbb{R}^p$, we use a function family $\bbPhi(\bbH, \bm{\theta})$ to model $\bbP(\bbH)$ and $\bbDelta(\bbH)$ as
\begin{equation} \label{eq_gnn}
\begin{split}
[\bbP(\bbH), \bbDelta(\bbH)] = \bbPhi(\bbH, \bm{\theta}).
\end{split}
\end{equation}
Substituting \eqref{eq_gnn} into \eqref{eq_problem111}, we translate it into a statistical learning problem. The goal here becomes to learn the optimal function $\bbPhi(\bbH, \bm{\theta}^*)$ with optimal parameters $\bbtheta^*$, which maximizes the objective function. This alternative learning problem can be solved without system models but only depending on system observations, as we demonstrate in Section \ref{primaldual}.

For the learning parameterization $\bbPhi(\bbH, \bm{\theta})$, we introduce the graph neural network (GNN). The GNN is well-know for its ability to exploit network structures to process data, which makes it a suitable candidate in our case. By interpreting RRHs and ANs as nodes and links between them as edges, the FSO fronthaul network can be abstracted as a weighted bipartite graph $\ccalG$. The graph representation matrix $\bbS$ captures the structure and the channel state information over network, which is defined as
\begin{equation}       
\bbS = \left(                
  \begin{array}{cc}   
    \bb0^{N \times M} & \bbH\\  
    \bbH^\top & \bb0^{M\times N}\\  
  \end{array}
\right).                 
\end{equation}
See Fig. \ref{fig_gnn} for details of the graph structure. The graph signal $\bbx = [x_1,\ldots,x_{N+M}]^\top$ is defined on the top of nodes, where $x_i$ is the signal value of $i$-th node indicating the status of RRH or AN.  

The key component constituting the GNN is the graph convolutional filter, which processes the graph signal $\bbx$ based on the graph matrix $\bbS$. Recall the graph shift operation $\bbS \bbx$ assigns to node $i$ the aggregated information $[\bbS \bbx]_i$ from its neighbors scaled by edge weights. It generalizes the signal shifting from the time domain to the graph domain and plays an important role in defining the graph convolution. Shifting $\bbx$ for $k$ times yields the signal $\bbS^k \bbx$, indicating $k$ node exchanges over graph. With a set of parameters $\theta_0,\ldots,\theta_K$, we aggregate $K$ shifted signals to obtain the higher-level feature
 \begin{equation} \label{eq_gf}
\begin{split}
\bbu = \sum_{k=0}^K \theta_k \bbS^k \bbx := \bbG(\bbS)\bbx
\end{split}
\end{equation}
where $\bbG(\bbS)$ is defined as the graph filter. Observe that $\bbG(\bbS)$ exploits the node information up to a neighborhood of radius $K$ and thus contains higher-level features generated from a more complete picture of the graph. If particularizing $\ccalG$ as the line graph where each node is a time instant and $\bbx$ is the signal sampled over time, the filter output $\bbG(\bbS)\bbx$ reduces to the traditional convolution. We can then think of \eqref{eq_gf} as the generalization of the convolution for graph signals.

The GNN is defined as a concatenation of layers, each of which consists of a bank of graph filters followed by a pointwise nonlinearity. To be more precise, at layer $\ell = 1,...,L$, we have $F$ input features $\{ x^{g}_{\ell-1} \}_{g=1}^{F}$. These features are processed by graph filters $\{ \bbG_{\ell}^{fg}(\bbS) \}_{fg}$ to produce higher-level features $\{ \bbu_{\ell}^{fg} \}_{fg}$ as in \eqref{eq_gf}. We aggregate the latter over the index $g$ and then apply the nonlinearity $\sigma(\cdot)$ to get the $f$-th output feature
\begin{equation} \label{eqn:GNN}
	\bbx_{\ell}^{f} \!=\! \sigma \left( \sum_{g=1}^{F} \bbu_\ell^{fg}\! \right)
		\!=\! \sigma \left( \sum_{g=1}^{F}\!\! \bbG_{\ell}^{fg}(\bbS) \bbx_{\ell\!-\!1}^{g} \!\right)\!,\forall f\!=\!1,...,F.
\end{equation}
The input of the GNN is the input of $1$-st layer $\bbx_0^1 = \bbx$ and the output of the GNN is the output of $L$-th layer $\bbx^1_L$. The learning parameters $\bbtheta$ are filter parameters $\{ \theta_{\ell 0}^{fg},\ldots,\theta_{\ell K}^{fg} \}_{\ell f g}$. We then denote 
\begin{equation} \label{eqn:GNN1}
\bbPhi(\bbx,\bbS,\bbtheta) = \bbx_L^1
\end{equation}
as the non-linear map of the GNN on the graph matrix $\bbS$ with parameters $\bbtheta$. The number of GNN parameters is independent of the network size, making it computationally efficient for
training compared with the dense deep neural network.

\subsection{Permutation Equivariance} 

With graph convolutional filters, the GNN accounts for the network structure in its parameterization. One key property obtained by doing this is the permutation equivariance, corresponding to that the resource allocation policy of \eqref{eq_problem111} is permutation equivariant. In particular, we define the permutation matrix $\bbPi$ as
\begin{equation} \label{eq_permutation}
\begin{split}
\bbPi \in \{ 0,1 \}^{(N+M) \times (N+M)}: \bm{\Pi} \bm{1} = \bm{1}, \quad \bbPi^\top \bm{1} = \bm{1}.
\end{split}
\end{equation}
Put simply, the permuted vector $\bbPi \bbx$ reorders the entries of $\bbx$ and the permuted matrix $\bbPi^\top \bbS \bbPi$ reorders the columns and rows of $\bbS$. We give the following theorem to formally state this property according to \cite{Gama2019}.
\begin{theorem}\label{theorem1}
Consider the GNN $\bbPhi(\bbx,\bbS,\bbtheta)$ with the graph signal $\bbx$, graph matrix $\bbS$ and parameters $\bbtheta$. For any permutation $\bbPi$, it holds that
\begin{equation} \label{eq_permutation2}
\begin{split}
\bbPhi(\bbPi \bbx,\bbPi\bbS \bbPi^\top ,\bbtheta) = \bbPi \bbPhi(\bbx,\bbS,\bbtheta).
\end{split}
\end{equation}
\end{theorem}
Theorem \ref{theorem1} establishes that the GNN with a permutation of underlying graph $\bbPi^\top \bbS \bbPi$ and input signal $\bbPi \bbx$ generates an equally permuted output. In the context of our case, it indicates that reordering RRHs or ANs in the FSO network indices an associated reordered resource allocation policy with policy parameters $\bbtheta$ unchanged. This property follows the intuition of resource allocation problems since the labelling of network nodes is arbitrary which should be reflected in our learning parameterization $\bbPhi(\bbx,\bbS,\bbtheta)$. We remark that the deep neural network does not satisfy the permutation equivariance without considering the graph structure in its architecture.

With the use of GNN, we represent \eqref{eq_gnn} as 
\begin{equation} \label{eq_gnn11}
\begin{split}
\!\bbPhi(\bbx,\!\bbS,\!\bbtheta)\!=\![\bbPhi_{\bbP}(\bbx,\!\bbS,\!\bbtheta),\! \bbPhi_{\bbDelta}(\bbx,\!\bbS,\!\bbtheta)]
\end{split}
\end{equation}
where $\bbPhi_{\bbP}(\bbx,\!\bbS,\!\bbtheta)=[\Phi_{\bbP 1}(\bbx,\!\bbS,\!\bbtheta),\ldots,\Phi_{\bbP K}(\bbx,\!\bbS,\!\bbtheta)]^\top=\bbP(\bbH)$ are assigned powers and $\bbPhi_{\bbDelta}(\bbx,\!\bbS,\!\bbtheta)=\bbDelta(\bbH)$ are selected ANs. With $\ccalP = [0,P_s]^N$ the space satisfying the peak power constraint \eqref{eq:safe} and $\Lambda$ the space restricting only one AN is selected by one RRH, we require $\bbtheta$ belongs to the set $\Theta = \{ \bbtheta | \bbPhi_{\bbP}(\bbx,\!\bbS,\!\bbtheta) \in \ccalP, \bbPhi_{\bbDelta}(\bbx,\!\bbS,\!\bbtheta) \in \Lambda \}$. The optimization problem \eqref{eq_problem111} is then translated into the following learning problem
\begin{alignat}{3} \label{eq_problem2}
\mathbb{P}:= &  \max_{\bbtheta \in \Theta} \ && \sum_{n=1}^N  \omega_n \sum_{m=1}^M \mathbb{E}\left[ C_{nm}(\bbH,\bbPhi(\bbx,\bbS,\bbtheta)) \right] ,             \\
        &  \st && \mathbb{E} \left[ \sum_{n=1}^N \Phi_{Pn}(\bbx,\bbS,\bbtheta) \right] \le P_t,   \nonumber \\
       &       &&\! \sum_{n\!=\!1}^N\! \mathbb{E}\left[ C_{nm}(\bbH,\bbPhi(\bbx,\bbS,\bbtheta)) \right] \!\le\! C_t,\forall m\!=\!1,...,M.   \nonumber
\end{alignat}
Note both the graph matrix $\bbS$ and the graph signal $\bbx$ are input variables of the GNN, which vary across time reflecting the instantaneous CSI of the FSO network and the status of RRHs and ANs. Our goal is to learn optimal GNN parameters $\bbtheta^* \in \Theta$ that maximize the objective and satisfy constraints.

\section{Primal Dual learning algorithm}\label{primaldual}

Consider the alternative problem \eqref{eq_problem2}. We develop a model-free primal-dual learning algorithm to train the graph neural network. With multiple constraints, it is straightforward to consider working in the dual domain. By introducing the non-negative dual variables $\bblambda = [\lambda_1,\ldots,\lambda_{M+1}] \in \mathbb{R}^{M+1}$, the Lagrangian of the problem is given by
\begin{align} \label{eq_lagran}
\mathcal{L}(\bbtheta,\bblambda) &= \sum_{n=1}^N  \omega_n \sum_{m=1}^M \mathbb{E}\left[ C_{nm}(\bbH,\bbPhi(\bbx,\bbS,\bbtheta)) \right] \nonumber\\
&+ \lambda_1 \left( P_t - \mathbb{E}\left[ \sum_{n=1}^N \Phi_{Pn}(\bbx,\bbS,\bbtheta)\right] \right) \\
&+ \sum_{m=1}^{M} \lambda_{m+1} \left( C_t - \sum_{n=1}^N\mathbb{E}\left[ C_{nm}(\bbH,\bbPhi(\bbx,\bbS,\bbtheta)) \right] \right)\!\nonumber
\end{align}
where constraints in \eqref{eq_problem2} are reinterpreted as weighted penalties in \eqref{eq_lagran}. The associated dual problem is to search for optimal $\bbtheta$ and $\bblambda$ that make a so-called ``saddle point'', i.e. respectively maximize and minimize the Lagrangian,
\begin{equation} \label{eq_dualprob}
\begin{split}
\mathbb{D} = \min_{\bblambda} \mathcal{D}(\lambda)=\min_{\bblambda} \max_{\bbtheta\in \Theta} \mathcal{L}(\bbPhi(\bbx,\bbS,\bbtheta),\bblambda).
\end{split}
\end{equation}
The null duality gap $\mathbb{D} - \mathbb{P}=0$ is achieved for convex optimization problems, in which we can work on the dual problem \eqref{eq_dualprob} instead without loss of optimality. The optimization problem \eqref{eq_problem2} is not necessarily convex with unknown system models, i.e., unknown objective function and constraints can be non-convex. However, for the learning parameterization $\bbPhi(\bbH, \bm{\theta})$ with strong function approximation ability, the duality gap of \eqref{eq_problem2} can be sufficiently small close to null \cite{Gao2019, Eisen2019}.

We develop a primal-dual learning algorithm to solve the dual poblem \eqref{eq_dualprob}, which then solves the primal problem \eqref{eq_problem2} as well. The primal-dual algorithm updates both the primal variables $\bbtheta$ and the dual variables $\bblambda$ iteratively with gradient descents, searching for the saddle point $(\bbtheta^*, \bblambda^*)$ that is maximal w.r.t. the primal variables and minimal w.r.t. the dual variables. Note that this saddle point is local because of the non-convexity, whose influence can be mitigated by methods such as performing algorithm multiple times to find the best solution. In particular, the algorithm is divided into two steps at each iteration:

(1) \emph{Primal step.} At $k$-th iteration, let $\bbtheta^k$ and $\bblambda^k$ be current primal variables and dual variables. We update the primal variables as
\begin{align} \label{eq_priup1}
&\bm{\theta}^{k+1} = \bm{\theta}^{k} + \delta^k \nabla_{\bm{\theta}} \mathcal{L}(\bm{\theta}^{k},\bblambda^{k}).
\end{align}
where $\delta^k>0$ is the step-size at iteration $k$. Observe that this update needs the system model to compute the gradients of the Lagrangian defined in \eqref{eq_lagran}. As we do not assume this to be available due to unknown system models, we resolve this issue using the policy gradient method common in reinforcement learning settings \cite{Sutton2000}. In particular, this approach considers the resource allocation policy $\bbPhi(\bbx,\bbS,\bbtheta)$ as samples drawn from a predetermined probability distribution $\pi_{\bbx,\bbS,\bm{\theta}}(\bbPhi)$ parameterized by the output of the GNN. With the use of the likelihood ratio identity, we can express the function gradient that takes
the form of $\nabla_{\bbtheta} \mathbb{E} [f(\bbH,\bbPhi(\bbx,\bbS,\bbtheta))]$ as 
\begin{equation} \label{eq_grad}
\begin{split}
\nabla_{\bbtheta} \mathbb{E} [f(\bbH,\!\bbPhi(\bbx,\!\bbS,\!\bbtheta))] \!=\! \mathbb{E}[f(\bbH,\!\bbPhi) \nabla_{\bm{\theta}}\! \log\! \pi_{\bbx,\bbS,\bm{\theta}}(\bbPhi)]
\end{split}
\end{equation}
where $\bbPhi$ is the random sample drawn from the distribution $\pi_{\bbx,\bbS,\bm{\theta}}(\bbPhi)$ that is determined by $(\bbx,\bbS,\bm{\theta})$. As such, we can represent the gradient of the Lagrangian in \eqref{eq_priup1} as
\begin{align} \label{eq_polgrad}
&\nabla_{\bm{\theta}} \mathcal{L}(\bm{\theta},\bblambda) \!\!=\!\! \mathbb{E}\! \left[\! \left(\! \sum_{n=1}^N  \!\!\omega_n \!\!\!\sum_{m=1}^M \! C_{nm}(\bbH,\bbPhi) \right.\!+\! \lambda_1^k\! \left(\!\! P_t \!-\!\! \sum_{n\!=\!1}^N \Phi_{Pn}\!\! \right) \right.\nonumber\\
&\left.\!+\! \sum_{m=1}^{M} \!\!\lambda_{m+1}^k\! \left(\! C_t\! -\! \sum_{n=1}^N\!C_{nm}(\bbH,\!\bbPhi)\!\right)\! \nabla_{\bm{\theta}}\! \log \pi_{\bbx,\bbS,\bm{\theta}}(\bbPhi) \right]\!\!
\end{align}
where $\mathbb{E}[\cdot]$ is approximated by sampling $N$ realizations from $\pi_{\bbx,\bbS,\bm{\theta}}(\bbPhi)$ and taking the average. It should be emphasized that \eqref{eq_polgrad} is model-free as the gradient of $\log \pi_{\bbx,\bbS,\bm{\theta}}(\bbPhi)$ can be computed given the distribution $\pi_{\bbx,\bbS,\bm{\theta}}(\bbPhi)$, while capacity values $C_{nm}(\bbH,\bbPhi)$ and $\Phi_{Pn}$ can be observed in the system. Furthermore, appropriate distributions shall be selected for $\pi_{\bbx,\bbS,\bm{\theta}}(\bbPhi)$ to satisfy the feasibility condition $\bbtheta \in \Theta$.

(2) \emph{Dual step.} With the obtained $\bm{\theta}^{k+1}$, the update of dual variables $\bblambda$ takes the form 
\begin{align} \label{eq_dualup1}
\lambda_1^{k+1} &=\left[ \lambda_1^{k} - \eta^k  \left( P_t -\mathbb{E} \left[ \sum_{n\!=\!1}^N\!\! \Phi_{Pn}(\bbx,\!\bbS,\!\bbtheta) \right] \right) \right]_+, \\
\lambda_{m+1}^{k+1} &\!\!=\!\!\left[\! \lambda_{m+1}^{k} \!-\! \eta^k \!\! \left(\!\! C_t \!-\!\mathbb{E}\! \left[ \sum_{n\!=\!1}^N\!\!C_{nm\!}(\bbH,\!\bbPhi(\bbx,\!\bbS,\!\bbtheta)\!)\! \right]\! \right)\! \right]_+ \label{eq_dualup2}
\end{align}
for all $m=1,\ldots,M$, where $\eta^k$ is the step-size and $[\cdot]_+$ is the non-negativity operator due to the definition of $\bblambda$. The dual update can be implemented with system observations $\Phi_{Pn}(\bbx,\bbS,\bbtheta)$ and $C_{nm}(\bbH,\bbPhi(\bbx,\bbS,\bbtheta))$, such that it is also model-free with no need of system models. As in the primal update, the expectation $\mathbb{E}[\cdot]$ can be approximated with the average of $S$ samples of $\bbH$.

The primal-dual update performed iteratively provides a model-free approach towards solving for the GNN parameters $\bbtheta$ and corresponding dual variables $\bblambda$ in problem \eqref{eq_problem2}. We summarize the whole training method in Algorithm 1.

{\linespread{0.9}
\begin{algorithm}[b] \begin{algorithmic}[1]
\STATE \textbf{Input:} Initial primal and dual variables $\bbtheta^0, \bblambda^0$
\FOR [main loop]{$k = 0,1,2,\hdots$}
      \STATE Draw CSI samples $\{ \bbH \}$ of batch size $S$, and compute the allocated resources $\{ \bbPhi \}$ according to the GNN $\bbPhi(\bbx,\bbS,\bbtheta)$ and the policy distribution $\pi_{\bbx,\bbS,\bm{\theta}^k}(\bbPhi)$
      \STATE Obtain channel capacity observations $C_{nm}(\bbx, \bbH,\!\bbPhi)$ with current allocated resources $\{ \!\bbPhi\! \}$ and the CSI $\{ \!\bbH\! \}$
      \STATE Compute the policy gradient $\nabla_{\bm{\theta}} \mathcal{L}(\bm{\theta}^k,\bblambda^k)$ by \eqref{eq_polgrad}
      \STATE Update the primal variable by \eqref{eq_priup1}  \\
	$\bm{\theta}^{k+1} = \bm{\theta}^{k} + \delta^k\nabla_{\bm{\theta}} \mathcal{L}(\bm{\theta}^k,\bblambda^k) \nonumber $
	\STATE Update the dual variable by \eqref{eq_dualup1}-\eqref{eq_dualup2}\\
	$\bblambda^{k+1} = \left[ \bblambda^{k} - \eta^k \nabla_{\bblambda} \mathcal{L}(\bm{\theta}^{k+1},\bblambda^k) \right]_+ \nonumber $
\ENDFOR

\end{algorithmic}
\caption{GNN primal-dual learning algorithm}\label{alg:learning} \end{algorithm}}


 \section{Simulation Results}\label{sec_numerical_results}

In this section, we present simulation results to corroborate our theory. We compare the GNN primal-dual learning policy with the baseline policy, i.e., equal power assignment and random AN selection, to show its strong performance.

\begin{figure}[t]
\centering
\includegraphics[width=0.47\linewidth, height=1\textheight, keepaspectratio]{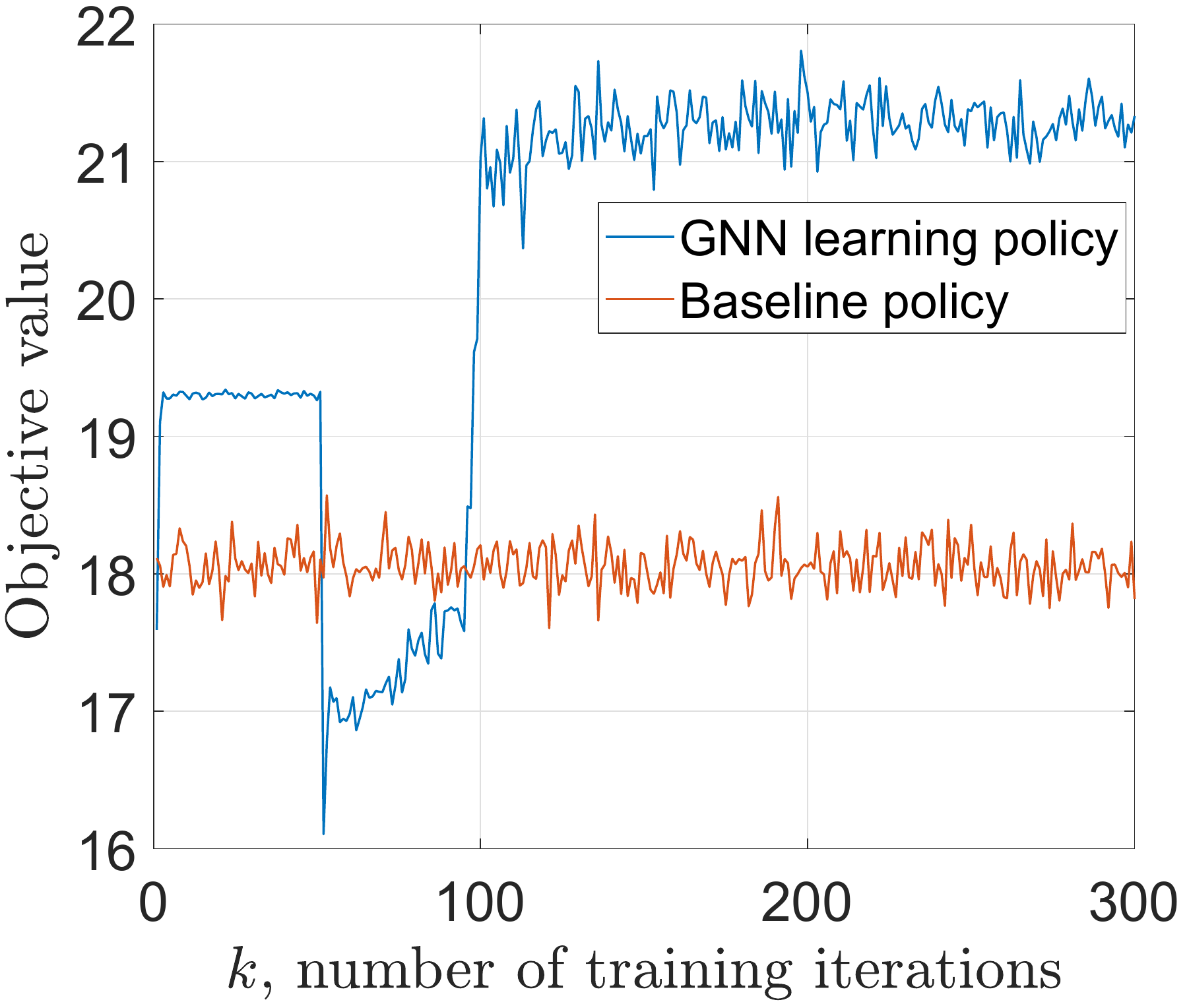}
\quad
\includegraphics[width=0.47\linewidth, height=1\textheight, keepaspectratio]{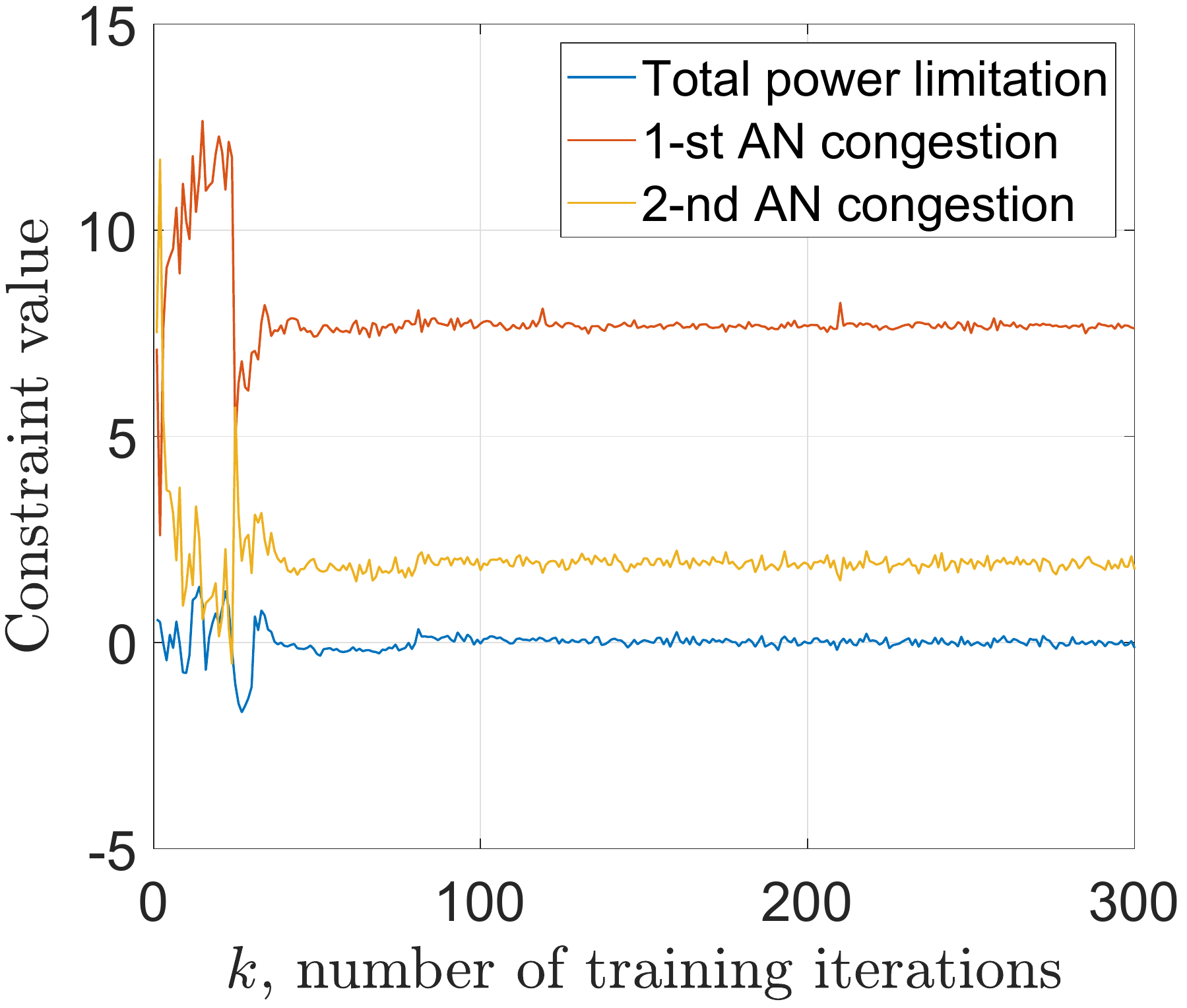}
\quad

\caption{The performance of the GNN policy and the baseline policy for $5$ RRHs and $2$ ANs with $P_t = 1.5 {\rm W}$ and $P_s = 0.5 {\rm W}$: (left) the objective value; (right) the constraint value.}\label{fig_simple_results}
\end{figure}

 \begin{figure}[t]
\centering
\includegraphics[width=0.47\linewidth, height=1\textheight, keepaspectratio]{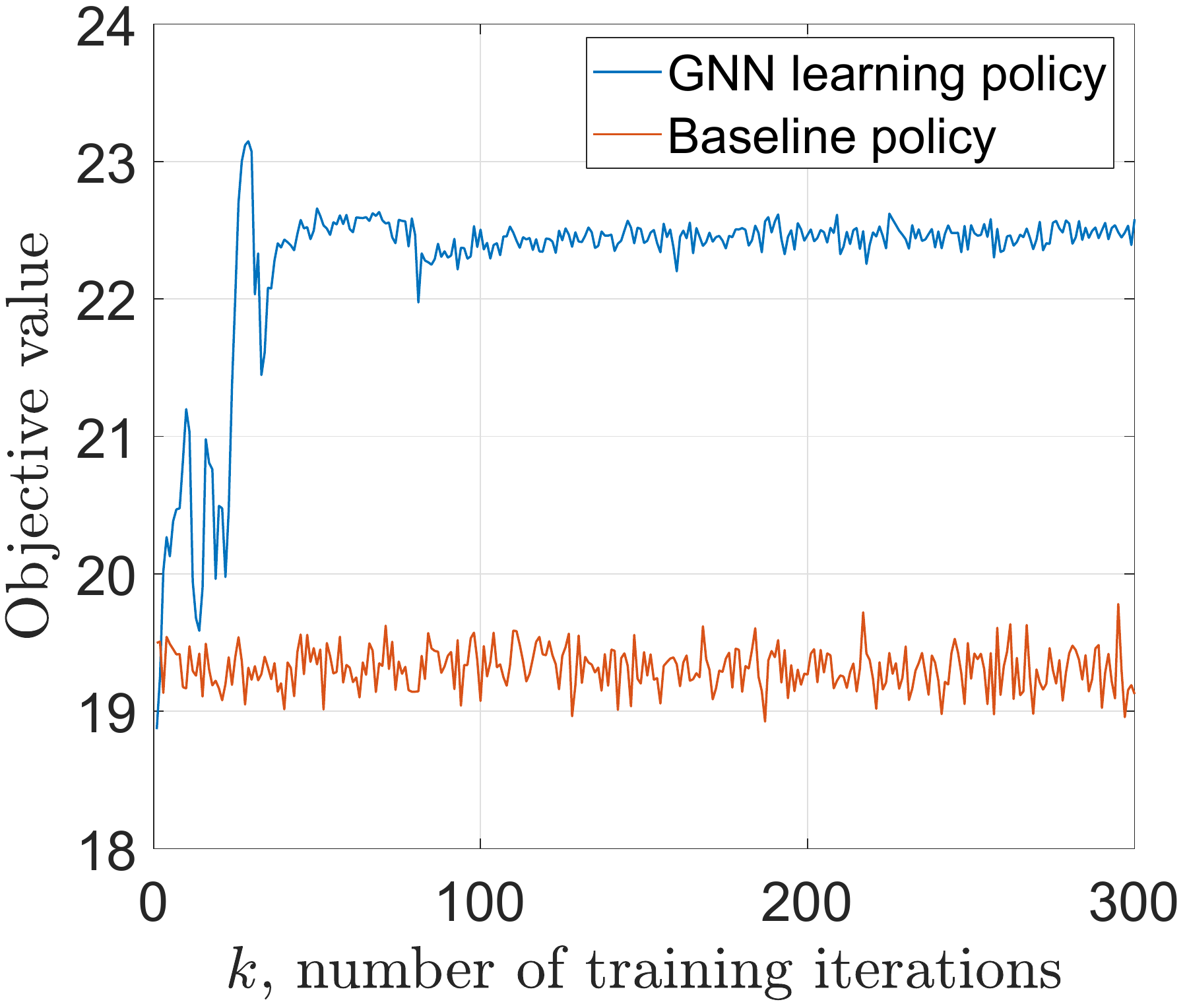}
\quad
\includegraphics[width=0.47\linewidth, height=1\textheight, keepaspectratio]{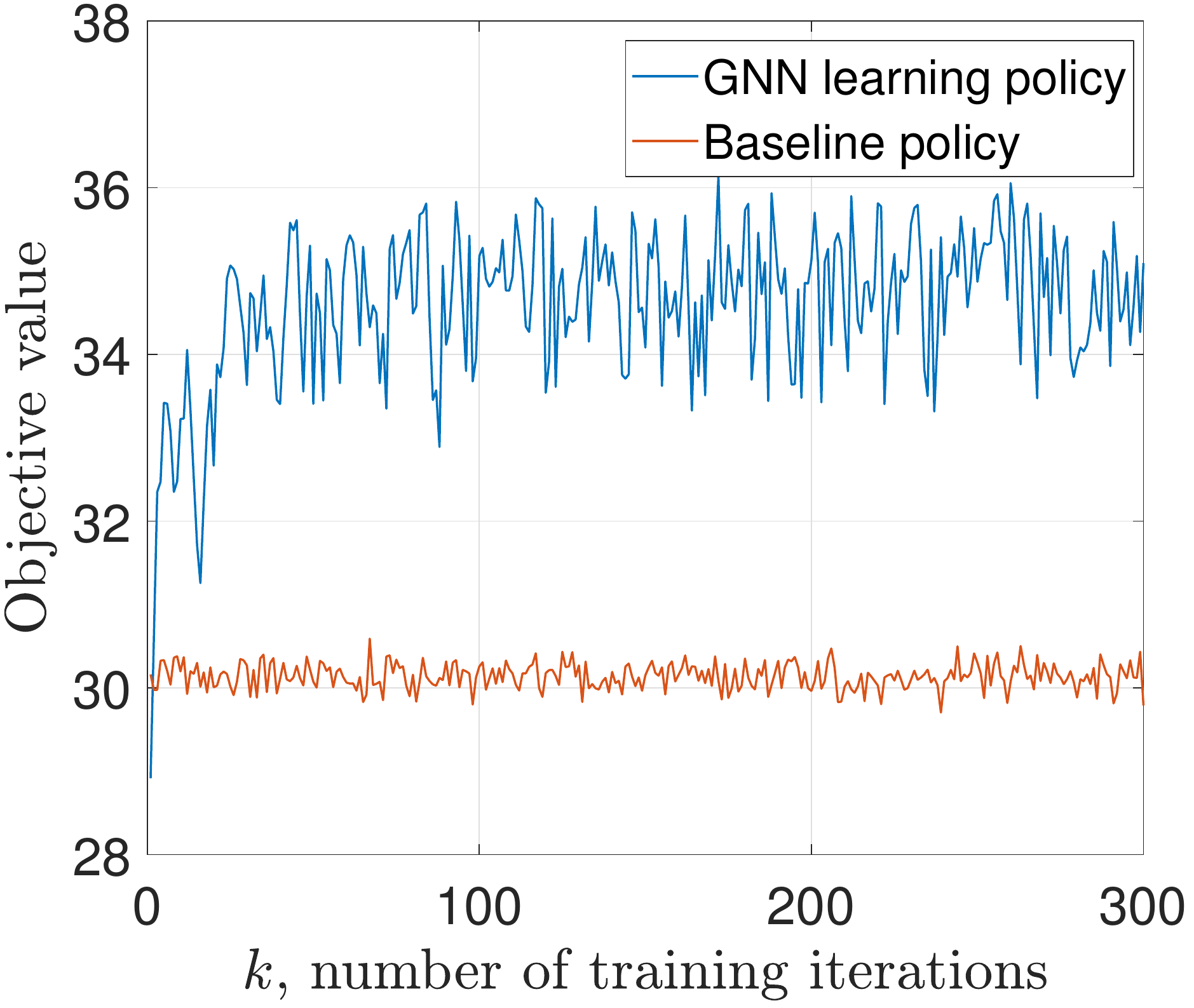}
\quad
\caption{The performance of the GNN policy and the baseline policy under different network scenarios: (left) power limitations with $P_t = 3 {\rm W}$ and $P_s = 1 {\rm W}$; (right) the network with $10$ RRHs and $4$ ANs.}\label{fig_simple_results}
\end{figure}

We consider a FSO fronthaul network as in Fig. 1. RRHs and ANs are distributed uniformly at random at locations $\bbr_n \in [-5{\rm km}, 5{\rm km}]^2$ and $\bba_m \in [-1{\rm km}, 1{\rm km}]^2$, and the weight vector $\bbomega$ is drawn randomly from zero to one. Since we are doing numerical simulations not physical experiments, the CSI samples $\{ \bbH \}$ and corresponding channel capacities $\{ C_{nm}(\bbx, \bbH,\!\bbPhi)\}$ cannot be observed. We use system models in \cite{Hassan2017} to compute these observation but keep in mind that the GNN learning policy works in a model-free manner. The GNN architecture is with $L=8$ layers, each of which contains $F=1$ graph filter of order $K=5$ followed by the ReLU nonlinearity $\sigma(\cdot) = [\cdot]_+$. The final layer passes through a sigmoid function to normalize the outputs and the latter is used as parameters of the policy distribution $\pi_{\bbx,\bbS,\bm{\theta}}(\bbPhi)$. The truncated Gaussian and categorical distributions are selected to satisfy the feasibility condition $\bbtheta \in \Theta$. The GNN is trained with the primal-dual learning algorithm, where the ADAM optimizer and geometrically decaying step-sizes are used for the primal update and the dual update, respectively. We point out this implementation is completely model-free, requiring only system observations $C_{nm}(\bbH,\bbPhi(\bbx,\bbS,\bm{\theta}))$ in practice. 

We first simulate on a small network with $N=5$ RRHs and $M=2$ ANs. The limitations are $P_t = 1.5 {\rm W}$, $P_s = 0.5 {\rm W}$ and $C_s = 20$. Fig. 3 (left) shows the performance, i.e., the weighted sum-capacity, achieved by the GNN learning policy and the baseline policy. We see that the learning process of the GNN converges as the training iteration increases. The GNN outperforms the baseline as we expected, and we emphasise that this performance improvement is obtained without explicit knowledge of capacity function models. The constraint values are shown in Fig. 3 (right). Both the power limitation and the data congestion constraints are satisfied as the learning process converges. This implies the feasibility of the optimal solution generated by the learned GNN. 

We then consider the GNN learning policy under different system scenarios; namely, larger power limitations $P_t = 3 {\rm W}$ and $P_s=1 {\rm W}$ (Fig. 4 (left)) and the larger network with $N=10$ and $M=4$ (Fig. 4 (right)). In general, the GNN maintains good performance for both cases. Specifically, we observe that the performance improvement of the GNN is emphasised compared with Fig. 3. This is because with larger allowed powers or at a larger network, the GNN gains more space to manipulate the resource allocation and thus better exhibits its learning capacity. Moreover, since the number of GNN parameters does not scale with the size of networks, either the training or the implementation of the GNN keeps computationally efficient for large networks. However, model-based algorithms will face a more complicated problem with more expensive computations.

We now evaluate the permutation equivariance of the GNN. For the FSO network with $N=5$ RRHs and $M=2$ ANs, we consider two permutations $\bbPi_1 \bbx = [x_3, x_4, x_5, x_2, x_1, x_6, x_7]^\top$ and $\bbPi_2 \bbx = [x_2, x_1, x_5, x_4, x_3, x_7, x_6]^\top$, i.e., relabelling RRHs and ANs. Table I shows the expected objective value over $100$ samples for the original and two permuted network scenarios. The results show that the same GNN learned from original network performs well for two permuted networks, verifying the permutation equivariance proposed in our theory. The small differences among three cases are because $100$ CSI samples are drawn from the probability distribution randomly.

\begin{table}[h] \label{table2}
\begin{center}
\caption{Performance of the GNN learning policy for permuted networks.}
\begin{tabular}{|l|l|l|l| p{2cm}|}
\hline
 & Objective value  \\ \hline
Original network &  21.397 \\  \hline
Network permutation $\bbPi_1$ & 21.405  \\ \hline
Network permutation $\bbPi_2$ & 21.381  \\ \hline
\end{tabular}  \vspace{-.55cm}
\end{center}
\end{table}


 \section{Conclusion} \label{sec_con}
We consider the optimal resource allocation in FSO fronthaul networks. The optimization problem takes the form of constrained statistical learning, where the resource allocation policy can be parameterized with the graph neural network. The GNN accounts for the network structure in its parameterization and thus exhibits the permutation equivariance, showing that it can achieve same performance on reordered FSO networks. We further develop a primal-dual learning algorithm to train the GNN, the implementation of which is model-free without requiring information of system models. This property is essentially important for FSO networks, in which cases sophisticated optical systems may be difficult to model or modelled inaccurately, leading to the performance degradation of model-based algorithms. Numerical simulations demonstrated the GNN is an effective parameterization for learning resource allocation policies and outperforms the baseline policy significantly. In the near future, we will extend the GNN learning algorithm to more FSO resource allocation scenarios.  

\bibliographystyle{IEEEbib}
\bibliography{FSO_learning,wireless_learning}

\begin{thebibliography}{10}

\bibitem{Gupta2015}
A.~Gupta and R.K. Jha,
\newblock ``A survey of 5g network: Architecture and emerging technologies,''
\newblock {\em IEEE Access}, vol. 3, pp. 1206--1232, 2015.

\bibitem{Wu2015}
J.~Wu, Z.~Zhang, Y.~Hong, and Y.~Wen,
\newblock ``Cloud radio access network (c-ran): a primer,''
\newblock {\em IEEE Network}, vol. 29, no. 1, pp. 35--41, 2015.

\bibitem{Poor2015}
M.~Peng, C.~Wang, V.~Lau, and H.~V. Poor,
\newblock ``Fronthaul-constrained cloud radio access networks: insights and
  challenges,''
\newblock {\em IEEE Wireless Communications}, vol. 22, no. 2, pp. 152--160,
  2015.

\bibitem{Alzenad2018}
M.~Alzenad, M.~Z. Shakir, H.~Yanikomeroglu, and M.~Alouini,
\newblock ``Fso-based vertical backhaul/fronthaul framework for 5g+ wireless
  networks,''
\newblock {\em IEEE Communications Magazine}, vol. 56, no. 1, pp. 218--224,
  2018.

\bibitem{Ahmed2018}
K.~Ahmed and S.~Hranilovic,
\newblock ``C-ran uplink optimization using mixed radio and fso fronthaul,''
\newblock {\em IEEE/OSA Journal of Optical Communications and Networking}, vol.
  10, no. 6, pp. 103--612, 2018.

\bibitem{Andrews2005}
L.~C. Andrews and R.~L. Phillips,
\newblock {\em Laser beam propagation through random media 2nd ed.},
\newblock Bellingham : SPIE Press, 2005.

\bibitem{Gao2017}
Z.~Gao, J.~Zhang, and A.~Dang,
\newblock ``Beam spread and wander of gaussian beam through anisotropic
  non-kolmogorov atmospheric turbulence for optical wireless communication,''
\newblock in {\em IEEE International Conference on Communications (ICC)
  Workshops}, 2017.

\bibitem{Gao2018}
Z.~Gao, Z.~Li, and A.~Dang,
\newblock ``Beam wander effects on scintillation theory of gaussian beam
  through anisotropic non-kolmogorov atmospheric turbulence for optical
  wireless communication,''
\newblock in {\em IEEE International Conference on Communications (ICC)
  Workshops}, 2018.

\bibitem{zhang2019ergodicity}
J.~Zhang, R.~Li, Z.~Gao, and A.~Dang,
\newblock ``Ergodicity of phase fluctuations for free-space optical link in
  atmospheric turbulence,''
\newblock {\em IEEE Photonics Technology Letters}, vol. 31, no. 5, pp.
  377--380, 2019.

\bibitem{Zhou2013}
H.~Zhou, D.~Hu, S.~Mao, and P.~Agrawal,
\newblock ``Joint relay selection and power allocation in cooperative fso
  networks,''
\newblock in {\em IEEE Global Communications Conference (GLOBECOM)}, 2013.

\bibitem{Zhou2015}
H.~Zhou, S.~Mao, and P.~Agrawal,
\newblock ``Optical power allocation for adaptive transmissions in
  wavelength-division multiplexing free space optical networks,''
\newblock {\em Digital Communications and Networks}, vol. 1, no. 3, pp.
  171--180, 2015.

\bibitem{Chatzidiamantis2013}
N.~D. Chatzidiamantis, D.~S. Michalopoulos, E.~E. Kriezis, G.~K. Karagiannidis,
  and R.~Schober,
\newblock ``Relay selection protocols for relay-assisted free-space optical
  systems,''
\newblock {\em IEEE/OSA Journal of Optical Communications and Networking}, vol.
  5, no. 1, pp. 92--103, 2013.

\bibitem{Hassan2018}
M.~Z. Hassan, M.~J. Hossain, J.~Cheng, and V.~C.~M. Leung,
\newblock ``Statistical delay-qos aware joint power allocation and relaying
  link selection for free space optics based fronthaul networks,''
\newblock {\em IEEE Transactions on Communications}, vol. 66, no. 3, pp.
  1124--1138, 2018.

\bibitem{Hassan2017}
M.~Z. Hassan, V.~C.~M. Leung, M.~J. Hossain, and J.~Cheng,
\newblock ``Delay-qos aware adaptive resource allocations for free space
  optical fronthaul networks,''
\newblock in {\em IEEE Global Communications Conference (GLOBECOM)}, 2017.

\bibitem{Sun2018}
H.~Sun, X.~Chen, Q.~Shi, M.~Hong, X.~Fu, and N.~D. Sidiropoulos,
\newblock ``Learning to optimize: Training deep neural networks for wireless
  resource management,''
\newblock {\em IEEE Transactions on Signal Processing}, vol. 66, no. 20, pp.
  5438--5453, 2018.

\bibitem{Xu2017}
Z.~Xu, Y.~Wang, J.~Tang, J.~Wang, and M.~C. Gursoy,
\newblock ``A deep reinforcement learning based framework for powerefficient
  resource allocation in cloud rans,''
\newblock in {\em IEEE International Conference on Communications (ICC)}, 2017.

\bibitem{Eisen2019}
M.~Eisen, C.~Zhang, L.~F.~O. Chamon, D.~D. Lee, and Ribeiro A.,
\newblock ``Learning to optimize: Training deep neural networks for wireless
  resource management,''
\newblock {\em IEEE Transactions on Signal Processing}, vol. 67, no. 10, pp.
  2775--2790, 2019.

\bibitem{Gao2019}
Z.~Gao, M.~Eisen, and A.~Ribeiro,
\newblock ``Optimal wdm power allocation via deep learning for radio on free
  space optics systems,''
\newblock in {\em IEEE Global Communications Conference (GLOBECOM)}, 2013.

\bibitem{Henaff2015}
M.~Henaff, J.~Bruna, and Y.~LeCun,
\newblock ``Deep convolutional network on graph-structured data,''
\newblock {\em arXiv preprint arXiv:1506.05163}, 2015.

\bibitem{Gama2019}
F.~Gama, A.~G. Marques, G.~Leus, and A.~Ribeiro,
\newblock ``Convolutional neural network architectures for signals supported on
  graphs,''
\newblock {\em IEEE Transactions on Signal Processing}, vol. 67, no. 4, pp.
  1034--1049, 2019.

\bibitem{Eisen2020}
M.~Eisen and A.~Ribeiro,
\newblock ``Optimal wireless resource allocation with random edge graph neural
  networks,''
\newblock {\em arXiv preprint arXiv:1909.01865}, 2019.

\bibitem{Zhan2020}
Z.~Gao, E.~Isufi, and A.~Ribeiro,
\newblock ``Stochastic graph neural networks,''
\newblock in {\em IEEE International Conference on Acoustics, Speech and Signal
  Processing (ICASSP)}, 2020.

\bibitem{Sutton2000}
R.~S. Sutton, D.~A. McAllester, S.~P. Singh, and Y.~Mansour,
\newblock ``Policy gradient methods for reinforcement learning with function
  approximation,''
\newblock in {\em Advances in Neural Information Processing Systems (NIPS)},
  2000.

\end{thebibliography}

\end{document}